\documentclass[conference]{IEEEtran}
\usepackage{latexsym}
\usepackage{amsfonts}
\usepackage{amssymb}
\usepackage{amsbsy}
\usepackage{amsmath}
\usepackage[dvips]{epsfig}
\usepackage{subfigure}

\ifCLASSINFOpdf
\else
\fi
\begin{document}
%
\title{Mobility Control for Machine-to-Machine LTE Systems}

\author{\IEEEauthorblockN{Beom Hee Lee, Seong-Lyun Kim}
\IEEEauthorblockA{School of Electrical and Electronic Engineering Yonsei University\\
134 Sinchon-Dong, Seodaemun-Gu, Seoul 120-749, Korea\\
Email:  bhlee@ramo.yonsei.ac.kr, slkim@yonsei.ac.kr} }




\maketitle

\begin{abstract}

In this paper, we propose an efficient mobility control algorithm for the downlink multi-cell orthogonal frequency division multiplexing access (OFDMA) system for co-channel interference reduction. It divides each cell into several areas. The mobile nodes in each area find their own optimal position according to their present location. Both the signal to interference plus noise ratio (SINR) and the capacity for each node are increased by the proposed mobility control algorithm. Simulation results say that, even the frequency reuse factor (FRF) is equal to 1, the average capacity is improved after applying the mobility control algorithm, compared to existing partial frequency reuse (PFR) scheme.

\end{abstract}




%
\IEEEpeerreviewmaketitle

\section{Introduction}

As the growing demand for wireless communication service, the cellular network system develops rapidly. To satisfy these demands, the long-term evolution (LTE) is introduced by 3rd Generation Partnership Project (3GPP) as the state-of-the-art technology. The object of the evolution is to obtain high date rates with low latency for human-to-human (H2H) communication. LTE Release 8 provides high peak data rates of 300 Mbps on the downlink and 75 Mbps on the uplink for a 20 MHz bandwidth \cite{Ghosh}.

Recently, not only the H2H communication but also machine-to-machine (M2M) communication is studied by standards bodies like ETSI \cite{ETSI} and 3GPP \cite{3GPP_b,3GPP_c}. The M2M communication is defined as a form of data communication between entities that do not necessarily need human interaction \cite{3GPP}. M2M communication can be used for smart metering, e-Health, city automation and automotive applications. For those use cases, the M2M devices are considered as mainly stationary machines or sensors installed in a fixed location.

However, some parts of the automotive application including unmanned V2V (vehicle to vehicle) and robotic networks will belong to the class of M2M with controllable mobility \cite{robot}, \cite{robo}. In this case, the M2M devices have mobility characteristics and we can control these mobility features (position or speed of each node) according to their application purpose. Even though M2M with controllable mobility is receiving less attention, controllable mobility itself is an interesting research dimension in wireless communications. In this paper, we evaluate the effect from controllable mobility nodes in cellular LTE networks with frequency reuse factor (FRF) 1 and partial frequency reuse (PFR) scheme \cite{Najjar}.

 OFDMA is the key technology for the future cellular wireless network. WiMAX uses OFDMA for IEEE 802.16 \cite{IEEE} while 3GPP LTE uses OFDMA on the downlink and Single-Carrier Frequency Division Multiple Access (SC-FDMA) on the uplink \cite{Ghosh}. In OFDMA, the whole spectrum is divided into a number of orthogonal sub-carriers to obtain high spectrum efficiency. Since the sub-carriers within a cell are orthogonal to each other, Inter-Cell Interference (ICI) mitigation becomes one of the most major concerns in LTE downlink. To avoid interference at cell edge, various fractional frequency reuse (FFR) schemes has been proposed \cite{Rahman}. Also, the two variations of the FFR scheme, PFR and soft frequency reuse (SFR) schemes are proposed \cite{Najjar}, \cite{Lei}, \cite {Huawei}. These schemes use a mix of FRF 1 and 3 to reduce the interference at cell edge region.

In this paper, we are interested in each node's controllable mobility to gain high spectral efficiency. Each node's proper movement guarantees higher SINR that naturally leads higher capacity. In this case, to define ``proper movement," we consider mainly two factors; SINR gain and lifetime of the nodes. To achieve higher SINR, each node tries to reduce the path loss from its own serving base station, while increasing the path loss between other cells. This means that, moving far away from other cells, the mobile node could move closer to its serving base station to increase SINR. The mobile node communicates and moves to other position by using its battery power which limits the lifetime. Considering these two factors - SINR gain and lifetime, we proposed a mobility control (MC) algorithm that allows nodes in the cell edge to achieve higher capacity.

The performance of the proposed MC algorithm is evaluated using system level simulation. The simulations focus on network-related issues like interference management and spectral efficiency. The SINR and the capacity for the cell edge node are investigated to see how proposed MC behaves in comparison with the existing PFR scheme.

 The rest of this paper is organized as follows: In Section II, system model is defined and described. Our mobility control algorithm is proposed in Section III. Section IV describes the simulation scenarios and settings. According to the simulation results, the capacity at the cell edge region is six times higher than the conventional scheme with FRF 1 and 1.5 times higher than PFR. Section V summarizes the key findings of this paper and discuss the effects of controllable mobility on M2M communications.

\section{System Model}

We investigate how our mobility control scheme enhances the performance the OFDMA cellular system adopting PFR \cite{Najjar}. We use a similar system model described in [8], as a general hexagonal cell structure with FRF 1.

\subsection{37 Multi-cell structure}
\begin{figure}[tbp]
\centering
\includegraphics[width=0.45\textwidth]{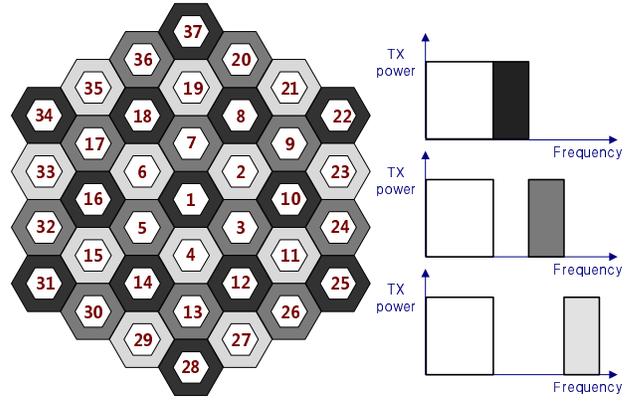}
\caption{37-cell-structure and frequency allocation}
\label{F:cell_structure}
\end{figure}

Consider the downlink of an OFDMA macro-cellular system with 37-cells (Fig. \ref{F:cell_structure}). In this system, to deploy PFR, each cell is divided into two parts; inner cell with FRF 1 and outer cell with FRF 3. In Fig. \ref{F:cell_structure}, the inner cell is illustrated as a white center cell in every single cell. The subcarriers allocated to the inner cells (mobile nodes) are reused in all 37 cells. However, the outer cell subcarriers are divided into three parts, which makes FRF 3. Then the total available subcarriers are classified into two groups corresponding to the area of inner cells and outer cells. We assume that nodes are distributed uniformly in each cell.

\subsection{The signal to interference plus noise ratio (SINR)}

We consider the distance-dependent path loss and shadow fading to determine the channel gain between a mobile node and a specific cell. The path loss model between node \begin{math} k \end{math} and serving cell \begin{math} m \end{math} can be formulated as,

\begin{equation}
PL(R) = 128.1 + 10 \alpha \log_{10}(R) + \theta_\sigma \quad(dB)
\label{eq:eqs1}
\end{equation}

\noindent where \begin{math} R \end{math} is the distance between node \begin{math} k \end{math} and serving cell \begin{math} m \end{math} in kilometer and the decay factor \begin{math} \alpha \end{math} is 3.76  \cite{Sawahashi}. Also, \begin{math} \theta_\sigma \end{math} is the lognormal shadowing effect which has a shadowing standard deviation \begin{math} \sigma \end{math}. Considering distance-dependant pass loss, the channel gain between node $ k $ and cell $ m $ on subcarrier $ i $ is described as follows:

\begin{equation}
G_{k,m,i} = 10^{PL(R)/10}
\label{eq:eqs2}
\end{equation}

The number of OFDM subcarriers allocated each node varies from less than one hundred to several thousands. In this subcarrier allocation, the orthogonality between nodes within a single cell is guaranteed. Hence, at least in principle, within one cell there is no intra-cell interference \cite{Parkvall}. The inter-cell interference is a considerable factor that influences the performance in terms of spectral efficiency. In this system, we indexed 37 cells and define the set $ S_m $ which contains interference cells for cell $ m $.

\begin{displaymath}
S_m = \{~j~|~j~\textrm{interference cells for cell $ m $}\}
\end{displaymath}

Without loss of generality, the SINR for the mobile node \begin{math} k \end{math} on subcarrier \begin{math} i \end{math} can be written as,
\begin{equation}
SINR_{k,i} = \frac{G_{k,m,i}P_{k,m,i}}{N_0 f_s + \sum_{l \in S_m}^{} G_{k,l,i}P_{k,l,i}}
\label{eq:eqs3}
\end{equation}

\noindent where $ P_{k,m,i} $ and $ G_{k,m,i} $ are respectively, the transmit power and the channel gain for node $ k $ on subcarrier $ i $ assigned to its serving cell. $ P_{k,l,i} $ and $ G_{k,l,i} $ are respectively, the transmit power and channel gain between the $ l^{th} $ co-channel cell and node $ k $ on subcarrier $ i $. $ N_0 $ is the power spectrum density of AWGN. $ f_s $ is the subcarrier spacing, with $ f_s = 1/T_s $, where $ T_s $ is the OFDM symbol period.

The nodes within the inner cell suffer from inter-cell interference from all other cells. For instance, node k within cell 1 receives interference from 36 other cells. On the other hands, the nodes in the outer region of cell 1 receives interference from only 12 cells.

\subsection{Subcarrier allocation}

Within a cell, we assume that all available subcarriers are allocated to mobile nodes fairly and dynamically. First of all, every node in the same cell obtains the same number of subcarriers. In this circumstance, since the number of subcarriers allocated to a node is dynamically changed according to the number of nodes in a cell, increasing the number of mobile nodes causes decreasing the capacity for a single node.

In case of PFR, the whole available subcarriers are dividend into two parts - one for inner cell nodes and the other for outer cell nodes. The subcarriers for outer cell nodes are separated into $ N $ parts, where $ N $ is the frequency reuse factor for the cell outer region. We assume that the mobile nodes are distributed uniformly. Thus, the number of subcarriers allocated to the inner cell and the outer cell is depending on how wide the area is. We represent $ R $ as the cell radius and $ \alpha R $ as the inner cell radius, where $ \alpha \in (0,1) $. Since the size of the inner cell and the outer cell are proportional $ \alpha ^2 R^2 \pi $ and $ (1-\alpha ^2 ) R^2 \pi $ respectively, the total subcarriers $ M $ is divided into $ \alpha ^2 M $ for the inner cell and $ (1- \alpha ^2 ) M $ for the outer cell.

\subsection{Reference model (PFR)}
 In [8], the optimal $\alpha$ say $\alpha_{opt}$, is determined by the average to variance ratio of the received SINR at nodes. That is, $\alpha_{opt}$ is obtained by solving a simple optimization problem given by,

\begin{equation}
\alpha_{opt} = \arg \max_{\alpha} \frac{\bar{\gamma}(\alpha)}{\sigma^{2}_{\gamma}(\alpha)}
\label{eq:eqs4}
\end{equation}

\noindent where $ \bar{\gamma}(\alpha) $ and $ \sigma^{2}_{\gamma}(\alpha) $ are the average and the variance of the instantaneous received SINR by nodes for a given $ \alpha $. In \cite{Najjar}, since the capacity analysis is skipped, we evaluate the capacity for each node at cell edge, in order to compare the performance on PFR and our MC scheme.

\subsection{Capacity}

The capacity for node $ k $ on subcarrier $ i $ can be expressed as \cite{Shen},

\begin{equation}
 C_{k,i} = f_s  log_2(1 + \beta SINR_{k,i})
\label{eq:eqs5}
\end{equation}

\noindent where, $\beta$ is the constant value with $ \beta = -1.5 / \ln(5 BER) $ and BER is set to $ 10^{-6} $ \cite{Lei}. The capacity for node $ k $, $C_k $, is obtained from summing the capacity for all subcarriers allocated to the node, which is written by,

\begin{equation}
C_k = \sum_i f_s  log_2 (1 + \beta SINR_{k,i})
\label{eq:eqs6}
\end{equation}

\noindent where, $ i $ denotes $i^{th}$ subcarrier allocated to node $ k $.

\section{Proposed Mobility Control Algorithm }

In our mobility control scheme, we assume that every node can estimate the distance to its serving BS and uses its residual energy for relocation of the position. Each node can also estimate its SINR gain and battery loss when location is changed. Considering the trade-off between SINR gain and battery loss, nodes choose their optimized moving distance. The cell edge nodes tend to move over longer distance to achieve high SINR, even if this will loose their battery power considerably.

\subsubsection{Lifetime Factor Function}
To consider each node's lifetime,
we define the \emph{lifetime factor} that has normalized values varying between  $0$ and $1$. At the initial
state, every node's lifetime factor is set to $1$, and it becomes
$0$ when the battery runs down. Other than $0$ and $1$, the lifetime
factor does necessarily not indicates the exact remaining
battery power. Let us now introduce a lifetime factor function that
relate lifetime factor values to the node's location in the
cell. Fig. 2 shows that $ k^{th} $ node's lifetime
factor $L_{k}(x)$ as a function of the moving distance $x$ to its serving BS. The shape of the function $L_{k}(x)$ differs, depending on the
initial position of the node. For example, in Fig. 2, one cell is divided into
four areas and nodes within each area have different lifetime factor
functions. For the cell edge node, the lifetime factor decreases
slowly at first, compared to the cell center node. Even if the
nodes within area $1$ and area $4$ move the same distance and use
exactly the same battery power, the nodes in area $4$ are aware that
their remaining lifetime is longer than the nodes within area $1$.
Cell edge nodes endure their battery loss more than cell center nodes.
However, if a node travels the maximum
moving distance $ x_{max} $, the lifetime factor goes
to $0$.


\begin{figure}[t]
\centering
\subfigure{
\includegraphics[width=0.2\textwidth]{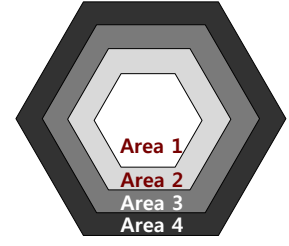}
\label{fig:sub1}
}
\subfigure{
\includegraphics[width=0.5\textwidth]{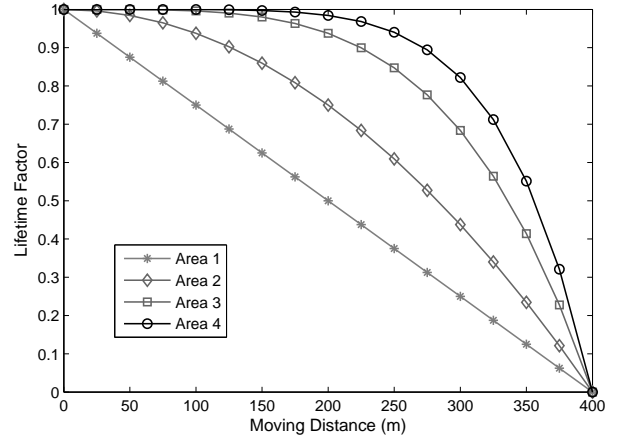}
\label{fig:sub2}
}
\label{F:lifetime}
\caption{Cell division and lifetime factor function ($x_{max} $ is set to 400 meters).}
\end{figure}

\subsubsection{Utility Function}

Every node calculates its utility from two factors: SINR gain and lifetime factor value. The value of the utility function (for node k) is expressed as,

\begin{equation}
U_{k}(x) = r_{k}(x)L_{k}(x)N_{k}(x)\quad \quad \  0 \leq x \leq x_{max}
\label{eq:eqs7}
\end{equation}

\noindent where $x$ is the moving distance for node $k$ and $ r_{k} (x) $, $ L_{k} (x) $ denote SINR gain and lifetime factor, respectively. The value of $ r(x) $ can be obtained from \eqref{eq:eqs3}. When the node moves over the distance $x$, the distance between the node and BSs in 37 cells are varied accordingly. For nodes within cell 1 in Fig. \ref{F:cell_structure}, in terms of SINR gain, the simplest and most effective way is to move toward the serving BS. $ N_{k}(x) $ represents the number of subcarriers allocated to node $ k $. $ N_{k}(x) $ could be $ \alpha_{opt}^2 M $ and $ (1- \alpha_{opt}^2 ) M $ according to the position of node, where M is the number of available subcarriers. The optimal moving distance $x_{opt}$ for node $k$ is obtained from the optimization problem described as follows:

 \begin{equation}
x_{opt} = \arg \max_{x} U_{k}(x)
\label{eq:eqs9}
\end{equation}

Each node decides its optimal moving distance to maximize the value of the utility function, which maximizes three factors: SINR level, lifetime factor function and the number of allocated subcarriers.

\section{Simulation Results}

 In this section, we present simulation results to evaluate the performance of the proposed MC scheme. As a comparison, we use PFR in a multi-cell OFDMA system (Fig. \ref{F:cell_structure}). The simulation parameters are chosen form \cite{Lei}, \cite{LTESPEC}. We assume that all the available subcarriers are transmitted with the same fixed power. The detailed simulation parameters are listed in TABLE \ref{T:simul para}.

Fig. 3 shows that the positions of nodes in cell 1. In this figure, the positions of nodes are classified into three cases: the initial position and changed position after applying MC under FRF 1 and PFR respectively. Since under the conventional scheme (FRF 1) the SINR is lower than PFR, Mobile nodes move longer distance under FRF 1 than PFR to achieve high capacity.

\begin{table}[tbp]
\centerline{
\begin{tabular}{|l|l|} \hline
    Parameter & Values \\ \hline
    Cellular layout & Hexagonal grid, 37 cell sites \\
    Cell radius (R) & 1000 m \\
    Distance dependent path loss & 128.1 + 37.6 log$_{10}$ (R) dB \\
    Shadowing standard deviation & 8 dB \\
    Channel bandwidth & 5 MHz \\
    Number of subcarriers & 300 \\
    Subcarrier spacing & 15 kHz \\
    BS transmit power & 43 dBm \\
    Noise density & -174 dBm/Hz  \\
    Max moving distance ($x_{max}$) & 400 m \\ \hline
\end{tabular}}
\caption{Simulation parameters} \label{T:simul para}
\end{table}

Fig. \ref{F:SINR} shows the downlink SINR level at each location in cell 1. At the cell edge, the interference from other cells is significantly increased. In our simulation, using the criterion (6), we obtain the optimal value of $ \alpha $, $ \alpha_{opt} = 0.467 $, which determines the radius of inner cell $ \alpha R $  as 467 (m). The SINR level of MC is in some part overlaped the performance of the conventional scheme (FRF = 1). This is because the frequency reuse factor of both schemes is equal to 1. However, since every node moves to the BS according to the equation \eqref{eq:eqs9} The SINR low bound of MC is higher than FRF 1. Due to the mobile movement to the BS, the SINR level and nodes do not exist in the interval from 667 (m) to 1000 (m).

\begin{figure}[tbp]
\centering
\subfigure{
\includegraphics[width=0.45\textwidth]{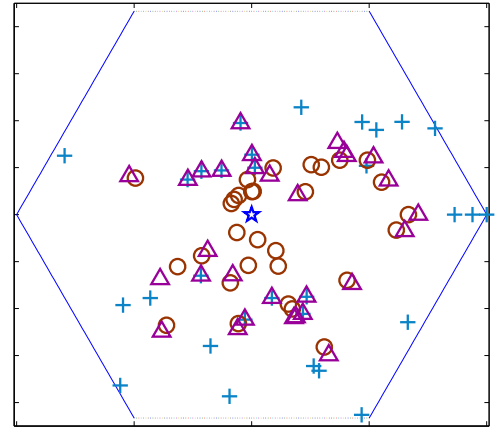}
\label{F:sub1}
}
\subfigure{
\includegraphics[width=0.19\textwidth]{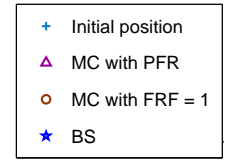}
\label{F:sub3}
}
\label{F:node}
\caption{position of 30 nodes in cell 1}
\end{figure}

\begin{figure}[tbp]
\centering
\includegraphics[width=0.5\textwidth]{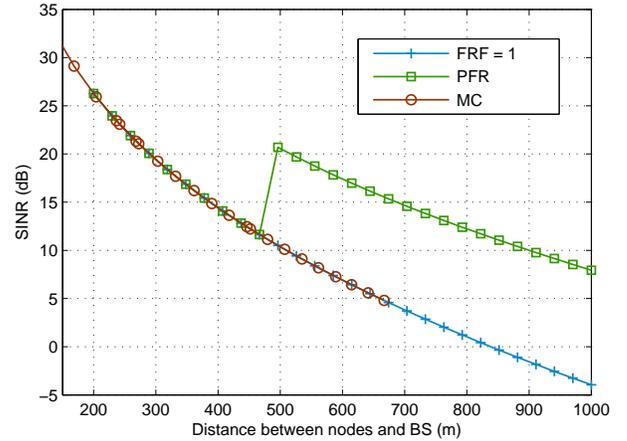}
\caption{Received SINR as a function of the distance}
\label{F:SINR}
\end{figure}

\begin{figure}[tbp]
\centering
\includegraphics[width=0.5\textwidth]{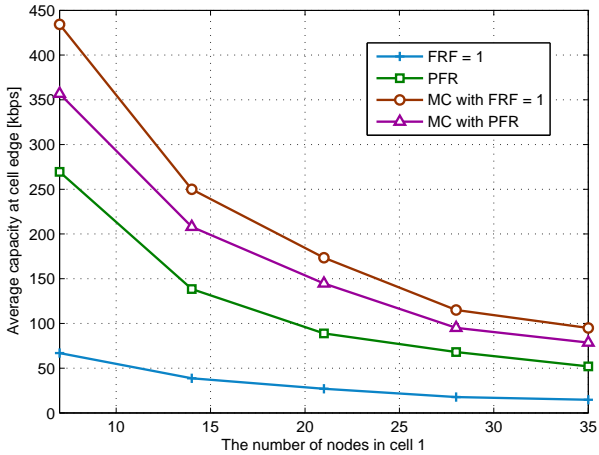}
\caption{Average capacity at cell edge region as a function of the number of nodes}
\label{F:capacity}
\end{figure}

Fig. \ref{F:capacity} depicts the average capacity at the cell edge region versus the total number of mobile nodes located in cell 1. We define the cell edge region as an area where the SINR level is lower than 0 (dB). In Fig. \ref{F:SINR}, we can see that SINR is almost 0 (dB) at 800 (m). At the cell edge region, the average capacity of PFR is higher than conventional scheme (FRF 1). However, the capacity improvement from PFR is limited, due to the fewer number of allocated subcarriers at cell edge. In Fig. \ref{F:capacity}, we notice that the average capacity of the nodes is increased after applying MC algorithm. The average capacity at the cell edge region is increased more than 6 times than FRF 1, and more than 1.5 times than PFR.

\begin{figure}[tbp]
\centering
\includegraphics[width=0.5\textwidth]{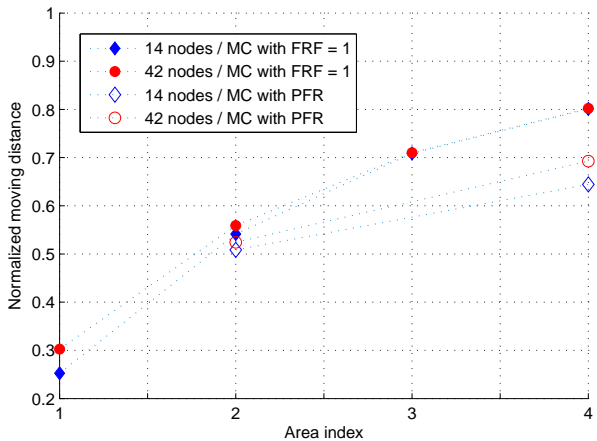}
\caption{Ratio of moving distance to the maximum moving distance at each area}
\label{F:LT_ratio}
\end{figure}

Fig. \ref{F:LT_ratio} shows that the average ratio of the moving distance of mobile nodes to the maximum allowable moving distance $x_{max} = 400$. Since we assume that the power of battery decreases linearly according to the moving distance, normalized moving distance implies energy consumption for each node. We also assumed that, in case of MC with FRF 1, cell 1 is divided into 4 areas and the lifetime factor function for each area is defined as shown in Fig. 2. For MC with PFR, the cell is divided into 2 areas; the inner cell and the outer cell. In this case, the lifetime factor functions for the inner cell and the outer cell are equivalent to those of area 2 and area 4 in Fig. 2. The optimal moving distance for each node within area 4 is much higher than the nodes within area 1. For the nodes in area 4, they use almost 80\% of their battery power to get higher SINR, while the nodes in area 1 use only 30\% (in case of MC with FRF 1). Fig. \ref{F:LT_ratio} also shows that nodes move longer distance under FRF 1 than PFR. Under FRF 1, since SINR level is lower than PFR, nodes consume more energy to achieve higher capacity. Especially, at area 4, there is a wide SINR gap between FRF 1 and PFR, nodes under FRF 1 use more energy and move longer distance than PFR. Thus, in terms of energy efficiency, nodes consume less energy when MC scheme is deployed under PFR. In Fig. \ref{F:LT_ratio}, we notice that the total number of nodes (14 and 42 nodes) has less relevance to the moving distance. Moving distances for nodes in a same area are similar each other regardless of the number of nodes in a cell.

\section{Conclusion}

Recently, M2M communication using cellular LTE network is studied actively \cite{3GPP_b}. If we can control the mobility for those M2M devices efficiently, the performance at the cell edge will be enhanced even under the worst frequency reuse planning (FRF is equal to 1). We analyzed how this controllable mobility affects the performance in a LTE cellular system and how much energy is consumed for the performance enhancement. In this paper, we proposed an efficient MC algorithm for multi-cell OFDMA systems. The simulation results showed the MC scheme brings higher channel capacity at the cell edge region, while minimizing the battery usage for each node. Numerical results showed that the proposed scheme provides better performance than the classical conventional scheme (FRF 1) and PFR proposed in \cite{Najjar}.

\section*{Acknowledgment}
The author's supported by Samsung Electronics Co., Ltd., Suwon 443-742, Korea.

\end{document}